\documentclass[a4paper]{article}

\usepackage{INTERSPEECH2022}
\usepackage{bbding}

\title{Robust End-to-end Speaker Diarization with Generic Neural Clustering}
\name{Chenyu Yang$^1$, Yu Wang$^1{^,}{^2}{^*}$ \thanks{${^*}$Corresponding author}}
\address{
  $^1$Cooperative Medianet Innovation Center, Shanghai Jiao Tong University \enspace$^2$Shanghai AI Laboratory
}
\email{\{yangcy,yuwangsjtu\}@sjtu.edu.cn}

\begin{document}

\maketitle
\begin{abstract}
 End-to-end speaker diarization approaches have shown exceptional performance over the traditional modular approaches. To further improve the performance of the end-to-end speaker diarization for real speech recordings, recently works have been proposed which integrate unsupervised clustering algorithms with the end-to-end neural diarization models. However, these methods have a number of drawbacks: 1) The unsupervised clustering algorithms cannot leverage the supervision from the available datasets; 2) The K-means-based unsupervised algorithms that are explored often suffer from the constraint violation problem; 3) There is unavoidable mismatch between the supervised training and the unsupervised inference. In this paper, a robust generic neural clustering approach is proposed that can be integrated with any chunk-level predictor to accomplish a fully supervised end-to-end speaker diarization model. Also, by leveraging the sequence modelling ability of a recurrent neural network, the proposed neural clustering approach can dynamically estimate the number of speakers during inference. Experimental show that when integrating an attractor-based chunk-level predictor, the proposed neural clustering approach can yield better Diarization Error Rate (DER) than the constrained K-means-based clustering approaches under the mismatched conditions. 

\end{abstract}

\section{Introduction}

Speaker diarization, which is often referred to as "who speaks when" problem, is an important speech processing task. The goal of a speaker diarization system is to estimate the temporal boundary of each talking speakers in real audio recordings~\cite{anguera2012speaker,park2022review}.
Traditional speaker diarization approaches first segment the audio streams into speech and non-speech regions, then the embeddings representing speaker characteristics are extracted from each speech frame. Popular used speaker embeddings include i-vectors~\cite{dehak2010front}, d-vectors~\cite{variani2014deep, heigold2016end}, and x-vectors~\cite{snyder2018x}. After the speaker embeddings are extracted, the embeddings belonging to the same speaker are clustered using unsupervised clustering methods to yield the diarization results, where agglomerative hierarchical clustering (AHC), K-means clustering, and spectral clustering are commonly used methods. Various speaker embeddings and clustering techniques have been explored for speaker diarization tasks in~\cite{sell2018diarization,diez2018but}. Although the traditional approaches can give good performance on various datasets, the main disadvantages of them are twofolds: first, in these approaches multiple components need to be optimized using separate criteria and second, it is difficult for them to deal with overlapped speech as the unsupervised clustering normally assumes that each segment can only be assigned to one speaker. Although recently some approaches are proposed leveraging the supervised neural clustering models to improve the diarization performance, such as the unbounded interleaved-state recurrent neural network (UIS-RNN)~\cite{zhang2019fully} and Discriminative Neural Clustering (DNC)~\cite{li2021discriminative} approaches, they still suffer from the aforementioned two problems.  

To solve those problems, end-to-end joint optimization-based models have been proposed which form the speaker diarization problem as a multi-label classification problem. The goal of these end-to-end framework is to achieve the diarization by a single neural network that is trained to directly optimize the diarization performance. The self-attentive EEND approach proposed in~\cite{fujita2019end} was optimized to calculate diarization results for every speaker in a mixture from input audio features using permutation invariant training (PIT)~\cite{fujita2019permutationfree}. EEND can only handle scenarios where numbers of speakers are fixed. It was later extended to deal with variable number of speakers by the end-to-end diarization model with encoder-decoder based attractor (EEND-EDA)~\cite{horiguchi2020end} and speaker-wise conditional EEND (SC-EEND)~\cite{fujita2020neural}. Although previous experiments have shown that these end-to-end models can have a better or similar performance as the traditional methods, they fail to handle some complicated situations, for example, long-form speech. In order to address these issues, some approaches have been proposed to combine the advantages of both end-to-end models and unsupervised clustering algorithms as a two-stage framework~~\cite{kinoshita2021integrating,kinoshita2021advances}. These two-stage approaches normally divide the complete audio sequence into chunks. Each chunk is processed by a neural-network-based model separately. An unsupervised clustering algorithm is then applied to determine the speaker correspondence among chunks. EEND-vector~\cite{kinoshita2021integrating,kinoshita2021advances} extends the EEND so that it can both diarize the subsequence and extract corresponding chunk-level speaker embeddings. These embeddings are clustered using the COP-K-means algorithm~\cite{wagstaff2001constrained}. This approach enables the original model to process long overlapped speech. Thus, it combines the advantages of both clustering and the EEND-based method. The approach proposed in~\cite{horiguchi2021towards} incorporates local attractors with the EEND-EDA model which splits the sequence after feeding it into the encoder and replaces the linear decoder with a local encoder-decoder-based attractor calculation module, in order to deal with the mismatch of the number of speakers between training and evaluation.

However, these methods use a constrained unsupervised clustering algorithms as the post-processing, which have some limitations: 1) Some constrained clustering algorithms like COP-K-means are not robust enough. They can not handle conflicts between constraints. These conflicts may be caused by incorrect estimation of the number of speakers. 2) The unsupervised clustering algorithms are only used in the inference stage. Although extra loss functions have been applied to make the training to be consistent with the inference, there is still unavoidable mismatch between the supervised training and unsupervised inference stages. This mismatch is especially obvious when estimating the number of clusters, which is required by most clustering algorithms. 

To mitigate the problems of using unsupervised clustering algorithms and achieve reliable diarization for real speech, in this paper we propose a generic supervised neural clustering algorithm for speaker diarization based on recurrent neural network. The proposed approach can leverage the advantages of both the Transformer-based end-to-end framework and the neural clustering approaches that have been explored in \cite{zhang2019fully,li2021discriminative}. First, the proposed neural clustering network can be jointly optimized with the front end processing and the embedding extraction modules, thereby accomplishing a fully supervised the end-to-end optimization and achieving more efficient use of the annotated data. Second, during inference it is able to dynamically estimate the number of speakers in the same way as training. Finally, the proposed neural clustering approach is generic in the sense that it can be integrated with any chunk-level predictor such as the EEND-EDA-based or EEND-vector models. 


\section{Proposed Framework: EEND with Nerual Clustering}

\subsection{Overview of two-stage frameworks}
A neural end-to-end diarization model receives a sequence of acoustic features $F\in\mathbb{R}^{D\times T}$ as the input, and outputs the active probabilities of each speaker $Y\in(0,1]^{S\times T}$, where $T$ is the length of input frames and $S$ is the number of speakers. A two-stage diarization framework normally consists of two parts: the chunk-level predictor processes the chunk-level subsequences and calculates local speaker representations called attractors. These attractors are then fed into the clustering module which concatenates the diarization result of each chunk with the speakers being reordered correctly. Such two stages are jointly optimized with a loss function $L$:
\begin{equation*}
  L = L_{pre} + L_{post}.
\end{equation*}

The proposed neural clustering approach can be integrated with any chunk-level predictor. Without loss of generality, in this paper we will focus on its integration with the the EEND-EDA predictor with local attractors~\cite{horiguchi2020end,horiguchi2021towards}. In the EEND-EDA framework, to deal with the mismatch of number of speakers, the input frames are first fed into a stack of Transformers.
And the output of encoder $E$ is then split into $N$ chunks, $E=\{E_n\}_{n=1}^{N}, E_n\in\mathbb{R}^{D\times L}$, where $L$ is the size of chunk. A chunk-level predictor is applied on each chunk separately.  The details of this predictor will be given in Section~\ref{sec:Sec 2.2}. After the chunk-level prediction stage, the proposed neural clustering module then re-concatenates the outputs together, which will be described in Section~\ref{sec:Sec 2.3}.

\subsection{Chunk-level diarization and attractor extraction}
\label{sec:Sec 2.2}
Similar to~\cite{horiguchi2021towards}, the chunk-level predictor is a local encoder-decoder-based attractor (EDA) calculator aiming to diarize subsequences and to calculate speaker attractors. it consists of an EDA module and a Transformer decoder layer.

The details of an EDA module can be found in \cite{horiguchi2020end}. It can be defined as:
\begin{equation*}
\hat{a}^n,\hat{z}^n  = {\rm EDA}(E_n),
\end{equation*}
where $\hat{a}^n\in\mathbb{R}^{D\times S_n}$ is the speaker representation and $\hat{z}^n\in(0,1]^{S_n}$ is the corresponding existence probabilities. The chunk-level active probability is computed by:
\begin{equation*}
  y^n = \sigma(E_n^T\cdot\hat{a}^n)\in(0,1]^{T\times S_n}.
\end{equation*}

 The chunk-level predictor is trained with two losses, namely the attractor existence loss and diarization loss. The attractor existence loss optimizes the estimation of each speaker's existence and the diarization loss optimizes the active probabilities with a permutation-invariant training method~\cite{yu2017permutation}:
\begin{align}
  L_{attr} &= \frac{1}{N}\sum_{n=1}^{N}{\rm BCE}(\hat{z}^n, z^n),~~~z^n = [\underbrace{1,\cdots,1}_{S_n},0],\\
  L_{diar} &= \min_{{\phi}\in\Phi}({\rm BCE}(y^n, l^n_{\phi})).\label{equ:pit}
\end{align}
Here $\Phi$ is the set of all possible permutations of $S_n$ speakers in the current chunk and $l^n_{\phi}\in \{0,1\}^{T\times S_n}$ is the corresponding diarization label. The total loss of chunk-level predictor are calculated as:
\begin{equation}
  L_{pre} = L_{diar} + L_{attr}.
\end{equation}

Since the local attractors are optimized to minimize the diarization error, they can not be used as the input for the neural clustering model directly. As in~\cite{horiguchi2021towards}, the local attractors, $\hat{a}^n$, are converted using a Transformer decoder before clustering:

\begin{equation}
  a^n = \mathrm{TransformerDec}(\hat{a}^n, E). \nonumber
\end{equation}

\subsection{RNN-based clustering methods}
\label{sec:Sec 2.3}
\begin{figure*}[t]
  \centering
  \includegraphics[width=\linewidth]{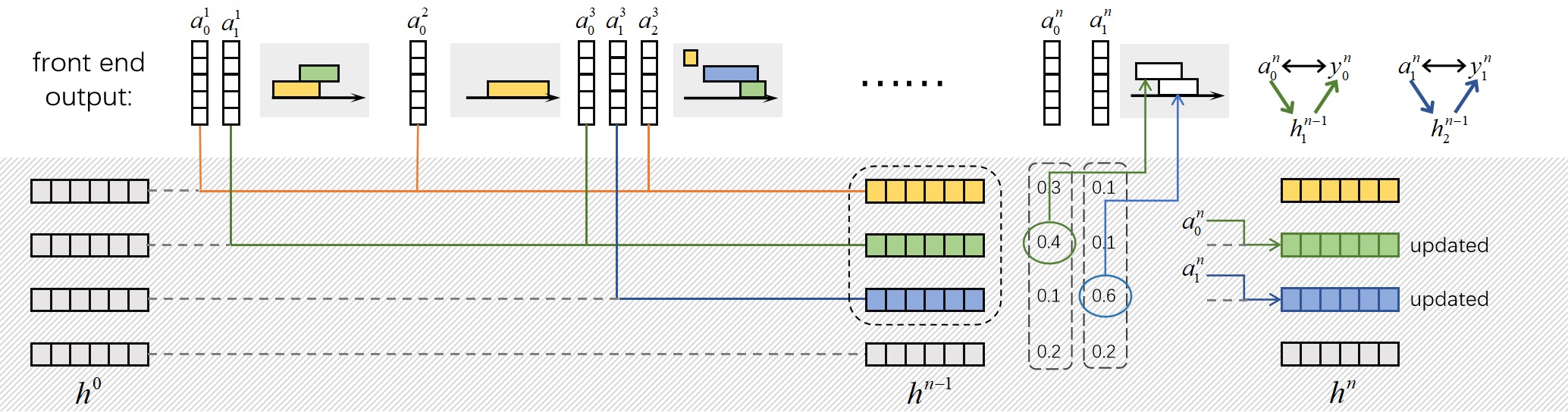}
  \caption{Diagram of the RNN-based sequential clustering method. Every color represents an individual speaker.}
  \label{fig:backend}
\end{figure*}

On the basis of the chunk-level predictor, we get pairs of attractors and diarization results for each chunk $\{(a^n_i, y^n_i)\}_{i=0}^{S_{n}-1}$. The goal of the neural clustering module is to reorder them by the optimal correspondence among chunks. 

In this paper, we explore a recurrent neural network (RNN) for the neural clustering because of its sequence modelling ability. Gated recurrent units (GRUs)~\cite{NIPS2017_f2fc9902} are leveraged for the clustering. As shown in Figure 2, the hidden states of the GRU cell can play a similar role as clusters in the unsupervised clustering algorithms~\cite{zhang2019fully}, which are used to model the global speakers. However, instead of being fixed during each iteration, these hidden states are updated adaptively according to the estimation of the attractors for each chunk. Thus, the estimation of the diarization outputs can be further split into two steps, namely the prediction and update steps. The details of these two steps are described as follows.

The prediction step aims to associate chunk-level attractors with the global hidden states, for which the RNN clustering (RC) model is optimized to estimate the probability $p^n_{i}=(p^n_{ij})_j\in(0,1]^{C}$ that the input attractor $a_i^n$ belongs to each speaker, which can be calculated as:

\begin{equation}
  p^n_{i} = \mathrm{softmax}\left((a_i^n)^Th^{n-1}\right), \nonumber
\end{equation}
where $h^{n-1}=(h^{n-1}_j)_j\in\mathbb{R}^{D\times C}$ are the previous hidden states. $C$ is the number of clusters present before. During the training phase, all hidden states have been initialized in the beginning, so $C$ is equalled to $S+1$.

Although the attractors are still out of order at this stage, we can link each attractor to a hidden state vector with the optimal permutation $\phi^n$ using Equation~(\ref{equ:pit}). Thus, the targets for training such a sequential clustering model are formed as:
\begin{align}
\label{equ:rij}
  r_{ij}^n & =\left\{\begin{array}{ll}
       1 &\phi^n_i = j\\  
       0 &otherwise \rlap{.}
  \end{array}
  \right. 
\end{align}
Because the optimal permutation has been estimated using Equation~(\ref{equ:pit}), permutation-invariant training is not required at this stage. Thus, a cross entropy (CE) loss is used to optimize the RNN clustering model:
\begin{equation*}
  L_{post} = \frac{1}{N}\sum_{n=1}^N\left(\frac{1}{S_n}\sum_{i=0}^{S_n-1}\sum_{j=0}^{C-1}-r^n_{ij}\cdot\log(p^n_{ij})\right).
\end{equation*}

After the prediction step, the goal of the update step is to adaptively adjust the hidden states at each step given the estimate of the attractors. Because $r_{ij}^n$ provides a mapping from the $i$th attractor, $a_i^n$, to the $j$th hidden state, $h_j^n$, the hidden states can be updated using a teacher-forcing strategy~\cite{williams1989learning}:
\vspace{0.1cm}
\begin{equation}
\label{equ: cluster}
  h_j^n=\left\{\begin{array}{ll}
       h_j^{n-1} &\forall i ~~~ r^n_{ij} = 0\\
       \mathrm{GRU}(a_i^n,h_j^{n-1}) &\exists i ~~~  r_{ij}^n = 1. 
  \end{array}
  \right.
\end{equation}
For the inference, the number of speakers is unknown. To dynamically estimate the attendance of new speakers, an initialized $h_0$ is concatenated with the previous states at every step. The optimal speaker permutation $\phi^n$, is estimated by:
$\phi^n=\arg\min_{\phi\in\Phi'}\prod_{i=0}^{S_n-1}p^n_{i,\phi_i}$,
where $\Phi'$ is the set of all possible permutations. It should be noticed that a cannot-link exists between every pair of attractors in the same chunk, therefore each hidden state is linked to no more than one attractors except for the last one. Then the hidden states are updated according to Equation~(\ref{equ:rij}) and (\ref{equ: cluster}).

After obtaining each $\phi^n$, 
the final output of the diarization is reordered by replacing the chunk-level output $y^n_i$ with $y^n_{\phi^n_i}$.


%

\section{Experiments}

\subsection{Data}
To evaluate the performance of the proposed RNN nerual clustering integrated EEND-
EDA approach, namely the EDA-RC approach, training and test datasets with different number of speakers were generated. The simulated datasets were generated using the data from the LibriSpeech\cite{panayotov2015librispeech} 360 hours of clean speech. The sampling frequency is 8kHz. A fixed number $n$ of speakers were first selected from this dataset, then for each speaker, utterances in the range of $[30/n, 60/n]$ were randomly selected. This simulating procedure was identical to the one that was introduced in \cite{fujita2019end}. It worth noting that the average length of silence intervals between utterances varied in order to maintain the overlap ratio in a reasonable range. The simulated training set includes 100,000 samples with 3 speakers. The details of the test sets for the simulation data, LS\_3, LS\_4 and LS\_5, are given in Table~\ref{tab:datasets}.

 The CALLHOME dataset, i.e., NIST SRE 2000 (LDC2001S97, Disk-8)~\cite{Phila2000nist},  was used to further evaluate the performance on the real data. This dataset has been widely used for speaker diarization experiments, which contains 500 sessions of multilingual telephonic speech. Although each session has 2 to 6 speakers, there are 2 dominant speakers in each conversation. For the experiments, it was split into training and test sets using the tools provided in the Kaldi Toolkit diarization recipe~\cite{povey2011kaldi}. Samples with $\leq 3$ speakers in the training set were picked to train the models. The test set was further split with various numbers of speakers in order to evaluate the performance of approaches for both matched and unmatched conditions.

\begin{table}[th]
  \caption{Summary of datasets that are used in the experiments}
  \label{tab:datasets}
  \centering
  \scalebox{0.9}{
  \begin{tabular}{lcccc}
  \toprule
  Dataset & Source & \#Spk & \#Sample & Overlap(\%) \\
  \midrule
  LS\_train & LibriSpeech & 3 & 100,000 & 49.6 \\
  LS\_3 & LibriSpeech & 3 & 500 & 42.7\\
  LS\_4 & LibriSpeech & 4 & 500 & 44.6 \\
  LS\_5 & LibriSpeech & 5 & 500 & 45.2\\
  \midrule
  CH\_ft & CALLHOME\_1 & 2-3 &  216 & 16.2 \\
  CH\_test & CALLHOME\_2 & 2-6 & 250 & 16.7 \\

  \bottomrule
  \end{tabular}}
\end{table}

\subsection{Experimental Setup}
We use 23-dimensional log-Mel-filterbanks with a 25-ms frame length and 10-ms frame shift as the input feature. Feature vectors were concatenated with their neighboring $[-7,7]$ frames and the concatenated features were subsampled by a factor of ten. Therefore, the total input dimension was 345 and the interval between frames was 100ms. This configuration was the same as that used in~\cite{fujita2019end}. The training samples were fed into models every 500 frames (50s) while the test samples remain un-split.

The baseline approaches include both one-stage and two-stage end-to-end approaches: the EEND approach~\cite{fujita2019end}, the EEND-EDA approach~\cite{horiguchi2020end} and the EEND-EDA with a COP-K-means unsupervised clustering on the local attractors (EDA+COP-K-means)~\cite{horiguchi2021towards}.  For the EEND and EEND-EDA approaches, a 4-layer Transformer with 256 hidden units were used as the encoder. Each layer had 4 heads and a dropout rate of 0.1. The end-to-end models were trained for 50 epochs with a batch size of 32. An Adam optimizer~\cite{kingma2015adam} with the same learning rate scheduler in~\cite{vaswani2017attention} was used during training.

For the EDA+COP-K-means and EDA-RC approaches, the chunk size was set to 50. Parameters of the EDA chunk-level predictor were the same as those in EEND-EDA. The models were trained for another 20 epochs by an Adam optimizer with a learning rate of $10^{-5}$.  Diarization Error Rate (DER) was used as the evaluation metric. Errors less than 0.25s around boundaries were tolerated. Please note that for the DER computation, all of the errors are evaluated including the overlapping speech segments.

\subsection{Experimental Results}

Table~\ref{tab:res1} shows the DERs for the Librispeech-based simulated datasets.  
Please note that the switching strategy described in \cite{horiguchi2021towards} is not adopted here, because the main purpose is to compare the unsupervised and supervised clustering methods instead of an overall performance. As expected, it can be seen that EEND and EEND-EDA approaches can yield performance for matched conditions but it gets worse with the increase of speakers. On the other hand, the EDA+COP-K-means approach takes an obvious advantage over the EEND-EDA approach and it outperforms the EEND-EDA by about 10\% DER under mismatched conditions. For our proposed EDA-RC approach, it can be seen that EDA-RC gives a DER reduction of 1.66\% and 1.56\% compared to the baseline approaches under mismatched conditions while the performance does not degrade much for the matched conditions.

Additionally, a refine method is also applied to the EDA-RC, where an extra decoding process was conducted with the hidden states initialized as the first-pass decoded hidden states. Table~\ref{tab:res1} shows that this strategy can yield small performance gains when the number of speakers is large. Finally, the performance of the EDA-RC with the oracle permutations (oracle) is also given as a reference. This shows the performance upper bound of the clustering integrated EEND approaches as the correct clustering methods are used to reorder the speakers of each chunk.


\begin{table}[h]
  \caption{Diarization Error Rate (\%) on simulated datasets. All models are trained with the 3-speaker training set.}
  \label{tab:res1}
  \centering
  \scalebox{0.9}{
  \begin{tabular}{ lcccc }
    \toprule
      & & match & \multicolumn{2}{c}{mismatch} \\
      \cmidrule(lr){3-3}\cmidrule(lr){4-5}
    Models & two-stage & LS\_3 & LS\_4 & LS\_5 \\ 
    \midrule
    EEND~\cite{fujita2019end} & \XSolidBrush & 5.19 & - & - \\
    EEND-EDA~\cite{horiguchi2020end} & \XSolidBrush & \textbf{4.93} & 33.79	& 41.82  \\
    + COP-K-means~\cite{horiguchi2021towards} & \checkmark & 6.93 & 22.48 & 32.36 \\
    \midrule
    EDA-RC (ours) & \checkmark & 5.17 & \textbf{20.82} & 30.80 \\
    EDA-RC+refine (ours) & \checkmark & 5.64 & 20.92 & \textbf{30.23}  \\
    \midrule
    EDA-RC (oracle) & \checkmark & 4.39 & 8.70 & 10.83 \\
    \bottomrule
  \end{tabular}}
\end{table}

\begin{table}[th]
  \caption{Diarization Error Rate (\%) on real dataset CALLHOME with different numbers of speakers.}
  \label{tab:res2}
  \centering
  \scalebox{0.9}{
  \begin{tabular}{lcccc}
    \toprule
      & \multicolumn{2}{c}{match} & \multicolumn{2}{c}{mismatch} \\
      \cmidrule(lr){2-3}\cmidrule(lr){4-5}
    Models & 2 & 3 & 4 & 5  \\ 
    \midrule
    EEND~\cite{fujita2019end} & 11.77 & 17.65 & - & -  \\
    EEND-EDA~\cite{horiguchi2020end} & \textbf{10.91} & \textbf{17.05} & 25.36 & 38.58   \\
    + COP-K-means~\cite{horiguchi2021towards} & 12.34 & 20.23 & 29.21 & 39.34  	\\
    ~~~ + switch\cite{horiguchi2021towards} & 11.51 & 18.51 & \textbf{25.19} & 39.01  	\\
    \midrule
    EDA-RC (ours) & 13.18 & 18.66 & 25.46 & \textbf{35.79}  \\
    ~~~ + switch & 11.78 & 18.20 & 26.57 & 37.04	\\
    \midrule
     EDA-RC (oracle) & 11.91 & 14.50 & 17.69 & 18.93  \\
    \bottomrule
  \end{tabular}}
\end{table}

In order to compare the performance on real audio recordings, models were also finetuned on CALLHOME. The beam size during decoding was set to 3. Models of the last $1/10$ epochs were averaged to obtain a better performance. 

Table~\ref{tab:res2} shows the DERs of various methods on the CALLHOME real datasets. It can be  the proposed EDA-RC approach yields 25.46\% and 35.79\% DER for the 4 and 5 speakers' conditions, which correspond to about 12.8\% and 9.0\% relative reduction compared to the EDA+COP-K-means approach. This validates the effectiveness and robustness of the proposed RNN-based clustering model. 

To make a more comprehensive comparison, a switching strategy that can dynamically decide whether to decode at chunk level or global level according to the potential number of speakers is also applied ~\cite{horiguchi2021towards} in this case. Thus, a global EDA loss is further added during the pretraining and finetuning of the EDA-RC model.

We notice that the gap between the clustering-integrated approaches and the conventional ones are rather small. This may be caused by the mismatch of data. As mentioned in Section~3.1, CALLHOME consists of real conversational speech where there are only two dominant speakers while all speakers have the same priority in the simulated data. Even though, with the switching method, clustering-integrated approaches can also have a better performance in some cases.

We further conduct ablation studies to analyze the performance of the proposed EDA-RC approach for various conditions. Considering an unsupervised clustering algorithm does not require a temporal order, it would be interesting to investigate the sensitivity of the RNN-based neural clustering to the order of the input chunks. The results are given in Table~\ref{tab:ablation} with various shuffled ratios of samples. It shows that DER decreased by about 1\% when the order is completely shuffled, which indicates the EDA-RC is not sensitive to the order of inputs. This may be resulted from the position independence of the Transformer encoder. In addition, Table~\ref{tab:ablation} shows that the impact of using a smaller decoding beam size on the performance of EDA-RC is also very small. 

\begin{table}[th]
  \caption{Ablation study}
  \label{tab:ablation}
  \centering
  \begin{tabular}{ ccccc }
    \toprule
      w/ $L_{global}$ & beam size & shuffled ratio &  \multicolumn{2}{c}{Test data} \\
        &  & $(\%)$ & LS\_3 & LS\_4 \\
    \midrule
        -  & 3 & 0 & 5.17 & 20.82 \\
        -  & 3 & 50 & 5.54 & 21.21 \\
        -  & 3 & 100 & 5.79 & 21.88 \\
        +  & 3 & 0 & 5.78 & 21.27 \\
        -  & 1 & 0 & 5.07 & 21.06 \\
    \bottomrule
  \end{tabular}
\end{table}

\section{Conclusions and Future Work}
In this paper, we propose a generic neural clustering method for two-stage end-to-end diarization models. We evaluate our model by solving the speaker mismatch problem. Compared to unsupervised clustering algorithms, our method is more robust and has a better performance. 
However, there is still some room for improvement. For example, such a method severely relies on the quality of local speaker representations, so a good chunk-level predictor is important. And it is sensitive to data adaptation, which should be solved in the future work. Apart from this, some advanced neural architectures like self-attention mechanism can also be used to improve the performance.


\bibliographystyle{IEEEtran}


\end{document}